\documentclass[a4paper,10pt]{article}
\usepackage[english]{babel}
\usepackage[T1]{fontenc}
\usepackage[utf8]{inputenc}
\usepackage{amssymb}
\usepackage{physics}
\usepackage{amsmath}
\usepackage{graphicx}
\usepackage[unicode]{hyperref}
\usepackage{tikz}
\usepackage{color}
\usepackage{genyoungtabtikz}
\usepackage[title]{appendix}
\usepackage{indentfirst}
\usepackage{bm}
\usepackage{xhfill}
\usepackage{bbm}
\usepackage{multirow}
\usepackage{array}
\usepackage[centertableaux]{ytableau}
\usepackage{multicol}
\usepackage[shortlabels]{enumitem}
\usepackage{forest}
\usepackage{float}

\hypersetup{
    colorlinks=true,
    linkcolor=blue,
    filecolor=magenta,      
    urlcolor=cyan,
}
 
\usepackage{amsthm}

\theoremstyle{definition}

\usepackage[nottoc]{tocbibind}

\usepackage{authblk}

\title{\textbf{Urn models, Markov chains and random walks in cosmological topologically massive gravity at the critical point}}

\author[1,2]{\textbf{Yannick Mvondo-She}}
\affil[1]{National Institute of Theoretical and Computational Sciences,  School of Physics and Mandelstam Institute for Theoretical Physics,
University of the Witwatersrand, Johannesburg, Wits, 2050, South Africa}
\affil[2]{DSI-NRF Centre of Excellence in Mathematical and Statistical Sciences (CoE-MaSS), South Africa}
\affil[ ]{\texttt{vondosh7@gmail.com}}
\affil[ ]{\texttt{yannick.mvondo-she@nithecs.ac.za}}

\date{}

\begin{document}

\maketitle

\begin{abstract}
We discuss a partition-valued stochastic process in the logarithmic sector of critical cosmological topologically massive gravity. By applying results obtained in our previous works, we first show that the logarithmic sector can be modelled as an urn scheme, with a conceptual view of the random process occurring in the theory as an evolutionary process whose dynamical state space is the urn content. The urn process is then identified as the celebrated Hoppe urn model. We next show a one-to-one correspondence between Hoppe's urn model and the genus-zero Feynman diagram expansion of the log sector in terms of rooted trees. In this context, the balls in the urn model are represented by nodes in the random tree model, and the "special" ball in this Pólya-like urn construction finds a nice interpretation as the root in the recursive tree model. Furthermore, a partition-valued Markov process in  which a sequence of partitions whose distribution is given by Hurwitz numbers is shown to be encoded in the log partition function. Given the bijection between the set of partitions of $n$ and the conjugacy classes of the symmetric group $S_n$, it is shown that the structure of the Markov chain consisting of a sample space that is also the set of permutations of $n$ elements, leads to a further description of the Markov chain in terms of a random walk on the symmetric group. From this perspective, a probabilistic interpretation of the logarithmic sector of the theory as a two-dimensional gauge theory on the $S_n$ group manifold is given. We suggest that a possible holographic dual to cosmological topologically massive gravity at the critical point could be a logarithmic conformal field theory that takes into account non-equilibrium phenomena.
\end{abstract}

\tableofcontents


\section{Introduction}
A special model of gravity in three dimensions has been the subject of considerable interest over the past fifteen years. Under the acronym CCTMG, critical cosmological topologically massive gravity has appeared to display remarkable properties, due to the fact that the Brown–Henneaux boundary conditions are relaxed \cite{Grumiller:2008qz}. In particular, within this framework, the theory has been shown to exhibit a new physical degree of freedom that behaves as a complex logarithmic function, the so-called logarithmic primary mode  

 \begin{eqnarray}
 \psi^{new}_{\mu \nu} := \lim_{\mu l \rightarrow 1}  \frac{\psi^M_{\mu \nu} (\mu l) - \psi^L_{\mu \nu}}{\mu l -1} = \left\{ -\ln \cosh{\rho} - i \tau \right\} \psi^L_{\mu \nu},
 \end{eqnarray}

\noindent which under the action of the $\rm{SL \left( 2, \mathbb{R} \right) \times SL \left( 2, \mathbb{R} \right)}$ AdS$_3$ isometry group generates descendant single particle logarithmic modes, and whose logarithmic branches extend the single particle class of solutions to a logarithmic multi-particle sector. 

The appearance of the new field brought a host of features into the theory. One exotic feature of this particular class of gravity is the non-unitarity of the theory, which arises through the emergence of Jordan cells. The latter being a defining property of logarithmic (L) CFTs \cite{Gurarie:1993xq,Flohr:2001zs,Gaberdiel:2001tr}, CCTMG was conjectured to be holographically dual to $c=0$ LCFTs, particularly effective in describing systems with (quenched) disorder \cite{Grumiller:2013at}, and was called log gravity \cite{Maloney:2009ck}.

A major result was obtained in the derivation of CCTMG's 1-loop partition function \cite{Gaberdiel:2010xv}, which was shown to agree with the partition function of an LCFT up to single particle. The original expression of the partition function reads

\begin{eqnarray}
\label{z tmg}
{Z_{\rm{CCTMG}}} (q, \bar{q})= \prod_{n=2}^{\infty} \frac{1}{|1-q^n|^2} \prod_{m=2}^{\infty} \prod_{\bar{m}=0}^{\infty} \frac{1}{1-q^m \bar{q}^{\bar{m}}}, \hspace{1cm} {\rm{with}} \hspace{0.25cm} q=e^{2i \pi \tau}. \bar{q}=e^{-2i \pi \bar{\tau}}.
\end{eqnarray}

\noindent From the identification of the first product as the three-dimensional gravity partition function $Z_{0,1}$ in \cite{Maloney:2007ud}, we have the convention 

\begin{eqnarray}
\label{z grav z log}
{Z_{\rm{CCTMG}}} (q, \bar{q})=  {Z_{\rm{gravity}}} (q, \bar{q}) \cdot {Z_{\rm{log}}} (q, \bar{q}),
\end{eqnarray}

\noindent with contributions 

\begin{eqnarray}
{Z_{\rm{gravity}}} (q, \bar{q})= \prod_{n=2}^{\infty} \frac{1}{|1-q^n|^2}, \hspace{0.5cm} \mbox{and}
\hspace{0.5cm} {Z_{\rm{log}}} (q, \bar{q}) = \prod_{m=2}^{\infty} \prod_{\bar{m}=0}^{\infty} \frac{1}{1-q^m \bar{q}^{\bar{m}}}.
\end{eqnarray}

\noindent Subsequent efforts to organize the multi-log excitations in a systematic way were undertaken, producing interesting results \cite{Mvondo-She:2018htn,Mvondo-She:2019vbx}. The log partition function was shown to be a $\tau$-function of the Kadomtsev–Petviashvili (KP) hierarchy \cite{Mvondo-She:2021joh}, expressed in terms of Schur polynomials $\chi_R (\mathcal{G}_1, \ldots,\mathcal{G}_n )$, which form a basis of the vector space of power series in variables with the coordinate sequence $\left( \mathcal{G}_k \right)_{k=1}^n$ as 

\begin{eqnarray}
\label{G_n}
\mathcal{G}_k \left( q,\bar{q} \right) = \frac{1}{|1-q^k|^2}, 
\end{eqnarray}

\noindent and which are parametrized by Young diagrams $R = \overbrace{\Yboxdim{4pt} \yng(1) \cdots \yng(1)}^n$ corresponding to the one-part partition $n^1 \vdash n$. The Schur polynomials playing a key role in the representation theory of the symmetric group, their appearance in the log partition function indicates the enumeration of product decomposition of permutations. The resulting expression of the log partition function as the generating function of the cycle decomposition of any permutation $\pi$ with exactly $j_k$ cycles of length $k$ in the symmetric group $S_n$ of all permutations on integers $1, \ldots , n$ is  

\begin{subequations}
\begin{align}
Z_{log} \left( \mathcal{G}_1, \ldots, \mathcal{G}_n  \right) &= 1 + \sum_{n=1}^{\infty} \chi_{\underbrace{\Yboxdim{4pt} \yng(1) \cdots \yng(1)}_n}  \cdot \left( q^2 \right)^n   \\
&= 1 + \sum_{n=1}^{\infty} \frac{1}{n!} Z_n \left( \mathcal{G}_1, \ldots, \mathcal{G}_n  \right) \left( q^2 \right)^n,
\end{align}    
\end{subequations}

\noindent with 

\begin{eqnarray}
Z_n \left( \mathcal{G}_1, \ldots, \mathcal{G}_n  \right) = \sum_{\pi \in S_n} \prod_{k=1}^n \mathcal{G}_k^{j_k}.   
\end{eqnarray}

\noindent Eventually, the logarithm of the log partition function is a solution of the KP I integrable hierarchy, signaling the presence of solitons unstable with respect to transverse perturbations in the theory.

 An interesting feature originally observed in CCTMG is the unstable aspect caused by the logarithmic mode, which renders the theory sensible. It was suggested in \cite{Grumiller:2008qz} that the instability could be an artifact of perturbation theory. This argument was shown to be encoded in the Hopf algebraic structure of the log partition function \cite{Mvondo-She:2022jnf}, which as a composition of functions, can be endowed with the coproduct of a Fa\`a di Bruno Hopf algebra expected to appear in any perturbation theory. 

 Generating functions that provide solutions to integrable hierarchies constitute a vast family which includes generating functions of Hurwitz numbers. The log partition function was shown to fall within that particular category of generating functions. Hurwitz numbers are an important class of combinatorial invariant which enumerate Riemann surfaces realized as branched covers of the Riemann sphere with specified branching structures. Let $h,g \in \mathbb{Z}$, $n \in \mathbb{N}$, $\lambda_1, \lambda_2, \ldots, \lambda_k$ partitions of $n$, and $f: X \mapsto Y$ be a holomorphic map of Riemann surfaces $X$ and $Y$ of genus $h$ and $g$, respectively. The Hurwitz number denoted
 
\begin{eqnarray}
H_{X \xrightarrow[]{n} Y} \left( \lambda_1, \lambda_2, \ldots, \lambda_k \right),    
\end{eqnarray}
 
 \noindent enumerates maps of Riemann surfaces classified as inequivalent (not necessarily connected) $n$-sheeted branched covering maps of the Riemann sphere, having $k$ branch points, with ramification profiles $\left( \lambda_1, \lambda_2, \ldots, \lambda_k \right)$, multiplied by the inverse of the order of the automorphism group of $f$ formed under composition by the collection of automorphisms of $f$. Equivalently, it is also $1/n!$ times the number of ways the identity element in the symmetric group on $n$ elements can be factorized into a product of $k$ elements belonging to specified conjugacy classes whose cycle lengths are given by the parts of the partitions $\left\{ \lambda_i \right\}_{i=1}^k$. In our case, we consider the disconnected Hurwitz number with partition sequence $\left\{ \lambda_i \right\}_{i=1}^2$, where $\lambda_1=\lambda_2$, and in particular $X=Y= \mathbb{CP}^1$ for which $h=g=0$. The log partition function becomes a generating function of Hurwitz numbers by expressing the one-part Schur polynomial as

 \begin{eqnarray}
 \chi_{\underbrace{\Yboxdim{4pt} \yng(1) \cdots \yng(1)}_n} = \sum_{\substack{\sum_{k=1}^n k j_k =n \\ n \geq 1\\j_k \geq 0 }} \left\{ H^{\bullet}_{0 \xrightarrow[]{n} 0} \left[ \left( [k]^{j_k} \right)_{k=1}^n,  \left( [k]^{j_k} \right)_{k=1}^n \right] \right\} \prod_{k=1}^n \mathcal{G}_k^{j_k},    
 \end{eqnarray}

\noindent with the disconnected Hurwitz numbers given in term of the sequence  $\left( [k]^{j_k} \right)_{k=1}^n = \left( [1]^{j_1}, \ldots, [n]^{j_n} \right)$  as

\begin{eqnarray}
\label{Hurwitz numbers}
H^{\bullet}_{0 \xrightarrow[]{n} 0} \left[ \left( [k]^{j_k} \right)_{k=1}^n,  \left( [k]^{j_k} \right)_{k=1}^n \right] = \prod_{k=1}^n \frac{1}{j_k!  (k)^{j_k}}, \hspace{1cm}  \sum_{k=1}^n k j_k=1,
\end{eqnarray}

\noindent where the sequence $\left( [k]^{j_k} \right)_{k=1}^n$ associated to the monomials $\prod_{k=1}^n \mathcal{G}_k^{j_k}$ are such that $[k]^{j_k}= \overbrace{k, \ldots. k}^{j_k ~ \rm{times}}$. 

A novel interpretation of the log partition function was established through the connection between its random combinatorial structures and a classical mutation model with numerous applications in mathematical population genetics, called the infinite-alleles model \cite{Mvondo-She:2023xel}. Such interpretation was possible by realizing that the Hurwitz numbers in the log partition function constitute a set of variables determined by a stochastic process, whose probability distribution is governed by a closed-form sampling formula for the infinite-alleles model called the Ewens sampling formula, with parameter $\theta=1$ in our case. The model is based on the fact that genes are taken as DNA sequences with a large number of sequence possibilities, the configuration of a sample of size $n$ is specified by an $n$-tuple ${\bf{j}}= \left( j_k \right)_{k=1}^n \vdash n$, where $j_k$ denotes the number of allelic types that appear exactly $k$ times in the sample, and mutations engender new alleles (gene types or DNA sequences) not already present in the existing population.

From the investigations done up to now, the log sector can be understood as an integrable model of random growth with a diffusion process on a singular manifold. In this model, solitons unstable with respect to modes of transverse perturbations are broken up in fragments of (defects) soliton clusters realizing the disordered landscape. In this work, building on the knowledge obtained, we make further probes into the log sector. 

In section 2, we shown that the log sector is a dynamical sample space that can be modelled using an urn scheme. An urn model can be described as a system of one or more urns containing balls of different colors \cite{johnson1977urn}. As a mathematical model, it represents a useful abstract tool for conceptualizing and modeling random processes through the action of drawing and returning rules, which control the evolution of the sample space of the urn (the contents of the urn). Of interest to us, a particular class of urn models called P\'{o}lya urn models consists of an initial single urn whose evolution follows a general method of ball drawing and replacement \cite{mahmoud2008polya}. We first show that the log sector can be described as an urn process whose evolution is built on the action of linear partial differential operators already constructed in \cite{Mvondo-She:2018htn} as a generator of the Heisenberg-Weyl algebra. The description is made more explicit by identifying the evolutionary stochastic process of the log sector to the celebrated Hoppe urn model \cite{hoppe1984polya}, a P\'{o}lya-like urn model whose construction was motivated by the idea of obtaining a simple and intuitive way to conceptualize the Ewens sampling formula. The mutation process described by the Ewens sampling formula is reproduced in the Hoppe urn model by introducing a special ball, the "mutator", which initiates the evolutionary process as the only ball in a single urn. Throughout the process, a ball is then withdrawn at random. If the mutator is withdrawn, it is placed back together with a ball of a new colour in the urn. If a ball of any other colour is drawn, one follows the mechanism of P\'{o}lya’s urn, and the selected ball is placed back in the urn together with another ball of the same color. The mutator ball is ignored in describing the urn configuration as it is always present with a prescribed $\theta$-weight as a device for generating mutations. A one-to-one correspondence is then shown between the stochastic dynamics of Hoppe’s urn model and genus-zero Feynman diagrams as a random recursive tree model. The balls in the urn are represented by nodes in the tree, and their urn decomposition into color clusters is represented by rooted trees whose branches have a corresponding color clustering of nodes.

In section 3, we show that a partition-valued Markov chain arises in the random clustering process describing the evolution of the log multiparticle sector with clusters counted in sequences of partitions. We demonstrate in our case, that the Markov property is implied by the property that the genus-zero double Hurwitz number generated by the log partition function, which enumerates the number of $\mathbb{CP}^1 \mapsto \mathbb{CP}^1$ branched covers with two branching points over zero and infinity with branching corresponding to two identical partitions of $n$, satisfies a recurrence relation. From the presence of the genus-zero double Hurwitz number which also describe the partition distribution of cycle lengths of random permutations on one hand, and the fact that Hopf algebras can fruitfully be used to encode the essential structure of Markov chains \cite{diaconis2014hopf} and random walks on group manifolds \cite{majid1993quantum,majid2000foundations} on the other hand, we then argue that the partition-valued stochastic process in the log sector can be further described as a special type of Markov chain, a random walk on a finite group, and in our particular case, a random walk on irreducible representations of $S_n$. This result leads to an interesting probabilistic interpretation of the log sector of CCTMG as a two-dimensional gauge theory on the symmetric group.

In section 4, we finally summarize our results, and give an outlook. In particular, our work shows evidence that the investigation of log gravity falls within the study of critical phenomena enhanced with out of equilibrium processes. As such we suggest that the holographic dual of CCTMG should be a logarithmic conformal field theory featuring non-equilibrium phenomena.

\section{Urn models and genus-zero Feynman diagrams}
In this section, from the information encoded in the log partition function, we show that the logarithmic sector of CCTMG can be constructed using an urn scheme. We first show that the urn model description can be realized algebraically via the Heisenberg–Weyl algebra. In our case, the latter was obtained in our previous work \cite{Mvondo-She:2018htn}. We then identify the correct urn model to be the celebrated Hoppe urn model, leading us to a concrete inference of the presence of defects in the theory as mentioned in \cite{Mvondo-She:2023xel}, as well as to a description of the Markov process associated to the Hoppe urn model using a recurrence relation on the Hurwitz numbers appearing in the log partition function. 

\subsection{From the log partition function to urn models}
A partition of a positive integer $n$ can be represented as a sum of positive integers $n= n_1 + n_2 + \ldots + n_m$, and be described equivalently as a sequence $n_{(1)} \geq n_{(2)} \geq \ldots  \geq n_{(m)}$, and in terms of a multiplicity sequence $\left( j_k \right)_{k=1}^n$ by defining $j_k = \# \{ i | \lambda_i=k \}$ as the number of times integer $k$ appears in the unordered set of so-called “occupancy numbers” $\left\{ n_1, \ldots, n_m \right\}$, such that $\sum_{i=1}^n j_i= m$ and $\sum_{i=1}^n ij_i= n$. These equivalent descriptions exhibit the well-known bijection between the set of partitions of $n$ and the conjugacy classes of the symmetric group $S_n$, established through the fact that the group $S_n$ can be divided into conjugacy classes according to a cycle structure specified
by the length $k$ of a cycle and the number $j_k$ of such a cycle. As a result, there is a one-to-one correspondence between an integer partition and a permutation $\pi \in S_n$ given by its cycle structure.

The log partition function generates counts of random permutations weighted by the number of cycles, where a permutation $\pi \in S_n$ of $\left\{ 1, \ldots, n \right\}$ is decomposed as a product of cycles with $\pi$ chosen uniformly with probability $1/n!$ and distributed according to the Ewens sampling formula, by encoding a permutation of $n$ with exactly $j_k$ cycles of length (or size) $k$ in the product 

\begin{eqnarray}
\label{prod}
\prod_{k=1}^n \mathcal{G}_k^{j_k}.    
\end{eqnarray}

The above information captured by Eq. (\ref{prod}) can be associated to the elements (balls) of an urn as follows. Consider an urn that consists of elements $\left\{ \textcolor{cyan}{\bullet}  ~ \textcolor{orange}{\bullet} ~ \textcolor{magenta}{\bullet} ~ \textcolor{cyan}{\bullet} ~ \textcolor{magenta}{\bullet} ~ \textcolor{cyan}{\bullet} ~ \textcolor{cyan}{\bullet} ~ \textcolor{orange}{\bullet} \right\}$ of length (or size) $n=8$. If we arrange the colors in clusters, and denote the clusters with ket notation as states, the states $\ket{\textcolor{cyan}{\bullet}   \textcolor{cyan}{\bullet}  \textcolor{cyan}{\bullet}  \textcolor{cyan}{\bullet}}$, $\ket{\textcolor{magenta}{\bullet} \textcolor{magenta}{\bullet}}$ and $\ket{\textcolor{orange}{\bullet} \textcolor{orange}{\bullet}}$ each appearing with multiplicity one can be encoded as $\mathcal{G}^{\textcolor{cyan}{1}}_{\textcolor{cyan}{\bullet} \textcolor{cyan}{\bullet} \textcolor{cyan}{\bullet} \textcolor{cyan}{\bullet}} \mathcal{G}^{\textcolor{magenta}{1}}_{\textcolor{magenta}{\bullet} \textcolor{magenta}{\bullet}} \mathcal{G}^{\textcolor{orange}{1}}_{\textcolor{orange}{\bullet} \textcolor{orange}{\bullet}}$. Finally, if we are only interested in the cluster size (i.e, the cycle length), we have the equivalence

\begin{eqnarray}
\mathcal{G}^{\textcolor{cyan}{1}}_{\textcolor{cyan}{\bullet} \textcolor{cyan}{\bullet} \textcolor{cyan}{\bullet} \textcolor{cyan}{\bullet}} \mathcal{G}^{\textcolor{magenta}{1}}_{\textcolor{magenta}{\bullet} \textcolor{magenta}{\bullet}} \mathcal{G}^{\textcolor{orange}{1}}_{\textcolor{orange}{\bullet} \textcolor{orange}{\bullet}}  \equiv \mathcal{G}^1_4 \mathcal{G}^1_2 \mathcal{G}^1_2 = \mathcal{G}^1_4 \mathcal{G}^2_2,   
\end{eqnarray}

\noindent where the partition of $n=8$ appears as $1 \cdot 4 + 2 \cdot 2$. This equivalence enables us to proceed with the description of the log sector as an urn model. 

\subsection{Urn models and Heisenberg algebra}
Enumeration problems in the probabilistic evolution of an urn process can be encoded in generating functions and represented in terms of polynomials and differential operators. Using an operator representation, a direct correspondence between the algebra of urn processes and the algebra of differential operators was established in \cite{blasiak2010urn}. At the centre of the correspondence is the Heisenberg–Weyl algebra. We use this formalism to show how the generic case of a single-mode urn model can be constructed in terms of appropriate differential operators.

We first consider an urn $\mathcal{U}_n$ containing $n$ balls. such that the content of the urn obey two elementary operations

\begin{center}
\begin{tabular}{ c l }
$\mathcal{X}:$ & putting a ball into the urn, i.e. $\mathcal{U}_n \mapsto \mathcal{U}_{n+1}$ \\
$\mathcal{D}:$ & withdrawing a ball from the urn, i.e. $\mathcal{U}_n \mapsto \mathcal{U}_{n-1}$   
\end{tabular}
\end{center}

\noindent We then represent the urn $\mathcal{U}_n$ by the polynomial $Z_n \left( \mathcal{G}_1, \ldots , \mathcal{G}_n \right)$, and the elementary operations $\mathcal{X}$ and $\mathcal{D}$ by multiplication $\hat{X}$ and derivative $\hat{D}$ operators respectively. If we consider the operators 

\begin{subequations}
\begin{align}
\hat{X} &= \mathcal{G}_1 + \sum_{k=1}^\infty k \mathcal{G}_{k+1}  \frac{\partial}{\partial \mathcal{G}_k}, \\
\hat{D} &= \frac{\partial}{\partial \mathcal{G}_1}
\end{align}
\end{subequations}

\noindent that we constructed in \cite{Mvondo-She:2018htn}, then the Heisenberg–Weyl algebra generated by the operators $\hat{X}$ and $\hat{D}$ satisfying the canonical commutation relation $\left[ \hat{D}, \hat{X} \right]=1$ manifests itself through the actions of multiplication and derivative operators on the polynomial $Z_n \left( \mathcal{G}_1, \ldots , \mathcal{G}_n \right)$ as 

\begin{subequations}
\begin{align}
\hat{X} Z_n&= Z_{n+1}, \\
\hat{D} Z_n &= n Z_{n-1}.
\end{align}
\end{subequations}

\noindent The interpretation of the correspondence between the algebra of urn processes and the algebra of differential operator both of which realize the Heisenberg–Weyl algebra is that, given an urn $\mathcal{U}_n$ containing $n$ balls, 

\begin{center}
\begin{tabular}{l}
$-$ there is only \textit{one} way of putting a ball into the urn $(\mathcal{X})$, \\
$-$ there are $n$ possible ways of withdrawing a ball from the urn $(\mathcal{D})$.  
\end{tabular}
\end{center}

\begin{figure}[h]
\begin{center} 
\begin{tikzpicture}
\filldraw[cyan] (-1.6,0) circle (0.1);
\filldraw[magenta] (-1.3,0) circle (0.1);
\filldraw[orange] (-1,0) circle (0.1);
\draw[->,thick] (0,0.2) -- (2,1);
\draw[->,thick] (0,-0.2) -- (2,-1);
\node [above] at (1,0.6) {$\hat{X}$};
\node [below] at (1,-0.6) {$\hat{D}$};
\filldraw[cyan] (2.5,1) circle (0.1);
\filldraw[magenta] (2.8,1) circle (0.1);
\filldraw[orange] (3.1,1) circle (0.1);
\filldraw[green] (3.4,1) circle (0.1);
\filldraw[cyan] (2.5,-1) circle (0.1);
\filldraw[magenta] (2.8,-1) circle (0.1);
\node [] at (4,-1) {OR};
\filldraw[cyan] (5.2,-1) circle (0.1);
\filldraw[orange] (5.5,-1) circle (0.1);
\node [] at (6.7,-1) {OR};
\filldraw[magenta] (7.9,-1) circle (0.1);
\filldraw[orange] (8.2,-1) circle (0.1);

\end{tikzpicture}    
\end{center}
\caption{Multiplication and derivation operations from an urn containing three balls.} \label{fig1}
\end{figure}

\noindent This can be illustrated by Fig. (\ref{fig1}). In what follows, we will see how the above urn model representation takes a well define description, in terms of the celebrated Hoppe urn model.

\subsection{Hoppe's urn model description of the log sector}
Hoppe's urn model was introduced for the first time in \cite{hoppe1984polya} within the framework of P\'{o}lya urn models, and with the distinctive feature that whereas P\'{o}lya urns scheme of ball drawing and replacement is defined via a finite set of admissible colors, the Hoppe urn model in contrast admits an infinite set of colors. 

Hoppe showed that the configuration of the colored balls after $n$ draws from the urn is distributed as the sampling formula of Ewens, that is, the distribution of the number of different gene types (alleles) and their frequencies at a selectively neutral locus under the infinitely-many-alleles model of mutation with rate $\theta > 0$. Hence, the following natural genetic interpretation for the Hoppe's urn scheme: new colors results from mutations and the black ball, which is ignored in describing the urn configuration is a device for introducing new mutations.

Hoppe's urn scheme was motivated by the biological phenomenon of the evolution of the partition of alleles according to the Ewens sampling formula, with new alleles represented as the aforementioned infinite set of colors arising via mutation realized by a fixed color. The evolutionary process starts with a single urn containing only the ball with the fixed "black" color, called the mutator as it engenders the mutation. As a ball is withdrawn at random, if the mutator is withdrawn, it is placed back together with a ball of a new colour in the urn. If a ball of any other colour is drawn, one follows the mechanism of Polya’s urn, and the selected ball is placed back in the urn together with another ball of the same color. The mutator ball is ignored in describing the urn configuration as it is always present with a prescribed weight $\theta=1$ and can be treated as a device that generates balls of new colors.

It is possible to show that this variant of Polya’s urn predicts a logarithmic, rather than power-law asymptotic increase of new colors in the urn \cite{hoppe1984polya,hoppe1987sampling}. The expected number of colors of ball in Hoppe’s urn after $n$ iterations can be computed explicitly \cite{mahmoud2008polya} as $\sum_{i=1}^n \frac{1}{i}$, i.e the $n$th partial sum of the harmonic series, which grows logarithmically in $n$. An alternative derivation for the expected number of different colors in the urn at level $n$ is now presented. For $n \in \mathbb{N}$, let $X_n$ denote the number of the color of the n-th ball added. Then, $\left( X_n \right)_{n \in N}$ is a stochastic process on $\mathbb{N}$ with $X_1 = 1$ and probabilities

\begin{eqnarray}
\label{proba}
\text{Pr}  \left[ X_{n+1} = k | X_n=x_n, \ldots, X_1=x_1  \right]  =
\begin{cases}
\frac{l_k(n)}{1+n},& k \leq l(n)\\
\frac{1}{1+n},& k=l(n)+1\\ 
0,& \text{otherwise},
\end{cases}
\end{eqnarray}

\noindent for every realization $x_1, \ldots,x_n$ of $X_1, \ldots,X_n$, $k \in \mathbb{N}$, where

\begin{eqnarray}
l(n) = \max{\{ d_i | 1\leq i \leq n \}}   \quad \text{and} \quad l_k(n)= \# \{ i | d_i=k, 1\leq i \leq n \},
\end{eqnarray}

\noindent with the probability of (color) coalescence as $\frac{l_k(n)}{1+n}$, and the probability of mutation, i.e the probability to draw a black ball and hence to add a new color in the urn as $\frac{1}{1+n}$, resulting in a rate of mutation that asymptotically vanishes. Now, let us define $D_n$ to be the number of different colors in the urn after $n$ draws. The random variable $D_n$ can be represented as the sum of the $n$ independent random variables $X_n$. Indeed, we place one extra ball in the urn and hence, after $i$ draws, there is a total of $1+i$ balls in the urn. The probability of picking a black ball in the $i$th draw is given by $1/i$, and the expected number of colors of ball in Hoppe’s urn after $n$ iterations is the sum $\sum_{i=1}^n \frac{1}{i}$, which viewed as a Riemann sum approximating an integral gives $\ln \left( n+1 \right)$ (see for instance \cite{durrett2008probability}, Eq. (1.10) page 17). Alternatively, this result can also be obtained as follows. The expected number of different colors $\mathbb{E} (D_n)$ after $n$ extractions follows the recurrence equation

\begin{eqnarray}
\mathbb{E} (D_{n+1})=\mathbb{E} (D_n) + \frac{1}{1+n},    
\end{eqnarray}

\noindent where the last term is the probability of extracting a brand new color given in Eq. (\ref{proba}). This recurrence equation is continuously approximated by

\begin{eqnarray}
\frac{d\mathbb{E} (D_n)}{dn}= \frac{1}{1+n}, \quad \mathbb{E} (D_0)=0,    
\end{eqnarray}

\noindent with solution 

\begin{eqnarray}
\mathbb{E} (D_n)= \ln \left( 1+n \right).    
\end{eqnarray}

The Hoppe's urn sequential construction scheme is illustrated up to order $n=3$ in Fig (\ref{fig2}), along with the multiplication operator action giving the algebraic correspondence of the probabilistic evolution of the urn process encoded in the polynomials of the log partition function.

\begin{figure}[h]
\begin{center} 
\begin{tikzpicture}
\filldraw[black] (0,0) circle (0.1);
\draw[->,thick] (0,-0.5) -- (0,-1.5);
\node [right] at (0,-1) {1};
\filldraw[black] (-0.15,-2) circle (0.1);
\filldraw[cyan] (+0.15,-2) circle (0.1);
\draw[->,thick] (-0.25,-2.5) -- (-1.25,-3.5);
\node [left] at (-0.8,-3) {$\frac{1}{2}$};

\draw[->,thick,cyan] (+0.25,-2.5) -- (+1.25,-3.5);
\node [right,cyan] at (+0.8,-3) {$\frac{1}{2}$};
\filldraw[black] (-1.8,-4) circle (0.1);
\filldraw[cyan] (-1.5,-4) circle (0.1);
\filldraw[magenta] (-1.2,-4) circle (0.1);

\filldraw[black] (+1.2,-4) circle (0.1);
\filldraw[cyan] (+1.5,-4) circle (0.1);
\filldraw[cyan] (+1.8,-4) circle (0.1);
\draw[->,thick] (-1.8,-4.5) -- (-4.75,-5.5);
\node [left] at (-4.0,-5) {$\frac{1}{3}$};

\draw[->,thick,cyan] (-1.5,-4.5) -- (-2.85,-5.5);
\node [left,cyan] at (-2.3,-5) {$\frac{1}{3}$};

\draw[->,thick,magenta] (-1.2,-4.5) -- (-1,-5.5);
\node [left,magenta] at (-1.1,-5) {$\frac{1}{3}$};

\draw[->,thick] (+1.2,-4.5) -- (+1,-5.5);
\node [right] at (+1.1,-5) {$\frac{1}{3}$};

\draw[->,thick,cyan] (+1.5,-4.5) -- (+2.85,-5.5);
\node [right,cyan] at (+2.3,-5) {$\frac{1}{3}$};

\draw[->,thick,cyan] (+1.8,-4.5) -- (+4.75,-5.5);
\node [right,cyan] at (+4.0,-5) {$\frac{1}{3}$};
\filldraw[black] (-5.2,-6) circle (0.1);
\filldraw[cyan] (-4.9,-6) circle (0.1);
\filldraw[magenta] (-4.6,-6) circle (0.1);
\filldraw[orange] (-4.3,-6) circle (0.1);

\filldraw[black] (-3.3,-6) circle (0.1);
\filldraw[cyan] (-3,-6) circle (0.1);
\filldraw[cyan] (-2.7,-6) circle (0.1);
\filldraw[magenta] (-2.4,-6) circle (0.1);

\filldraw[black] (-1.4,-6) circle (0.1);
\filldraw[cyan] (-1.1,-6) circle (0.1);
\filldraw[magenta] (-0.8,-6) circle (0.1);
\filldraw[magenta] (-0.5,-6) circle (0.1);

\filldraw[black] (+0.5,-6) circle (0.1);
\filldraw[cyan] (+0.8,-6) circle (0.1);
\filldraw[cyan] (+1.1,-6) circle (0.1);
\filldraw[orange] (+1.4,-6) circle (0.1);

\filldraw[black] (+2.4,-6) circle (0.1);
\filldraw[cyan] (+2.7,-6) circle (0.1);
\filldraw[cyan] (+3,-6) circle (0.1);
\filldraw[cyan] (+3.3,-6) circle (0.1);

\filldraw[black] (+4.3,-6) circle (0.1);
\filldraw[cyan] (+4.6,-6) circle (0.1);
\filldraw[cyan] (+4.9,-6) circle (0.1);
\filldraw[cyan] (+5.2,-6) circle (0.1);
\node [] at (+6.2,0) {1};
\draw[->,thick] (+6.2,-0.5) -- (+6.2,-1.5);
\node [right] at (+6.2,-1) {$\hat{X}$};
\node [] at (+6.2,-2) {$Z_1$};

\draw[->,thick] (+6.2,-2.5) -- (+6.2,-3.5);
\node [] at (+6.2,-4) {$Z_2$};
\node [right] at (+6.2,-3) {$\hat{X}$};

\draw[->,thick] (+6.2,-4.5) -- (+6.2,-5.5);
\node [] at (+6.2,-6) {$Z_3$};
\node [right] at (+6.2,-5) {$\hat{X}$};
\end{tikzpicture}    
\end{center}
\caption{Log sector evolution in terms of sample paths and multiplication operation $\hat{X}$ (i.e, adding a ball) in Hoppe’s urn up to order $n=3$.}
\label{fig2}
\end{figure}

\noindent We conclude this section by giving a genus-zero Feynman diagram interpretation of Hoppe's urn model from the latter's description of the log sector. A family of random tree model associated to the Hoppe’s urn model has appeared in the literature under the name of Hoppe tree \cite{leckey2013asymptotic}. In the special case $\theta=1$ which is ours, this tree model is also known as the random recursive tree model, in which the recursive trees grow in a similar manner as permutations in cycle notation. Given the structure of the log partition function which can be understood in terms of the distribution of cycle counts according to the $\theta$-biased permutation with parameter $\theta=1$, the content of the urns in Hoppe's scheme corresponds to the graphical representation of the log partition function in terms of genus-zero Feynman diagrams \cite{Mvondo-She:2022jnf}. The $n$-level Feynman diagram dynamics form a randomly growing rooted tree model whose growth parallels the evolution of Hoppe's urn model.

We previously argued \cite{Mvondo-She:2023xel} that in the log partition function expansion over rooted trees, solitonic clusters are pinned at the roots which represent the pinning sites. The interesting observation here is that, as the balls in the Hoppe urns are represented by nodes in the Hoppe trees, the mutator ball that initiates the Hoppe's urn model corresponds to the root of the Hoppe tree, responsible for the diffusion of the fragmented clusters that realize the disorder landscape (see Fig. \ref{fig3}). 

\begin{figure}[h]
\begin{center} 
\begin{tikzpicture}
\filldraw[black] (0,0) circle (0.1);
\draw[->,thick] (0,-0.5) -- (0,-1.5);
\node [right] at (0,-1) {1};
\filldraw[black] (0,-2.25) circle (0.1);
\draw[-] (0,-1.75) -- (0,-2.25);
\filldraw[cyan] (0,-1.75) circle (0.1);
\draw[->,thick] (-0.25,-2.5) -- (-1.25,-3.5);
\node [left] at (-0.8,-3) {$\frac{1}{2}$};

\draw[->,thick,cyan] (+0.25,-2.5) -- (+1.25,-3.5);
\node [right,cyan] at (+0.8,-3) {$\frac{1}{2}$};
\filldraw[cyan] (-1.8,-3.75) circle (0.1);
\draw[-] (-1.5,-4.25) -- (-1.75,-3.85);
\filldraw[black] (-1.5,-4.25) circle (0.1);
\draw[-] (-1.5,-4.25) -- (-1.25,-3.85);
\filldraw[magenta] (-1.2,-3.75) circle (0.1);

\filldraw[black] (+1.5,-4.25) circle (0.1);
\draw[-] (+1.5,-4.25) -- (+1.5,-3.75);
\filldraw[cyan] (+1.5,-3.75) circle (0.1);
\draw[-] (+1.5,-3.65) -- (+1.5,-3.25);
\filldraw[cyan] (+1.5,-3.25) circle (0.1);
\draw[->,thick] (-1.8,-4.5) -- (-4.75,-5.5);
\node [left] at (-4.0,-5) {$\frac{1}{3}$};

\draw[->,thick,cyan] (-1.5,-4.5) -- (-2.85,-5.5);
\node [left,cyan] at (-2.3,-5) {$\frac{1}{3}$};

\draw[->,thick,magenta] (-1.2,-4.5) -- (-1,-5.5);
\node [left,magenta] at (-1.1,-5) {$\frac{1}{3}$};

\draw[->,thick] (+1.2,-4.5) -- (+1,-5.5);
\node [right] at (+1.1,-5) {$\frac{1}{3}$};

\draw[->,thick,cyan] (+1.5,-4.5) -- (+2.85,-5.5);
\node [right,cyan] at (+2.3,-5) {$\frac{1}{3}$};

\draw[->,thick,cyan] (+1.8,-4.5) -- (+4.75,-5.5);
\node [right,cyan] at (+4.0,-5) {$\frac{1}{3}$};
\filldraw[black] (-4.75,-7.5) circle (0.1);
\draw[-] (-4.75,-7.5) -- (-5.15,-7);
\filldraw[cyan] (-5.15,-7) circle (0.1);
\draw[-] (-4.75,-7.5) -- (-4.75,-7);
\filldraw[magenta] (-4.75,-7) circle (0.1);
\draw[-] (-4.75,-7.5) -- (-4.35,-7);
\filldraw[orange] (-4.35,-7) circle (0.1);

\filldraw[black] (-2.85,-7.5) circle (0.1);
\draw[-] (-2.85,-7.5) -- (-3.15,-7);
\filldraw[magenta] (-3.15,-7) circle (0.1);
\draw[-] (-2.85,-7.5) -- (-2.55,-7);
\filldraw[cyan] (-2.55,-7) circle (0.1);
\draw[-] (-2.55,-6.9) -- (-2.55,-6.5);
\filldraw[cyan] (-2.55,-6.5) circle (0.1);

\filldraw[black] (-1.0,-7.5) circle (0.1);
\draw[-] (-1.0,-7.5) -- (-1.3,-7);
\filldraw[cyan] (-1.3,-7) circle (0.1);
\draw[-] (-1.0,-7.5) -- (-0.7,-7);
\filldraw[magenta] (-0.7,-7) circle (0.1);
\draw[-] (-0.7,-6.9) -- (-0.7,-6.5);
\filldraw[magenta] (-0.7,-6.5) circle (0.1);

\filldraw[black] (+1.0,-7.5) circle (0.1);
\draw[-] (+1.0,-7.5) -- (+0.7,-7);
\filldraw[orange] (+0.7,-7) circle (0.1);
\draw[-] (+1.0,-7.5) -- (+1.3,-7);
\filldraw[cyan] (+1.3,-7) circle (0.1);
\draw[-] (+1.3,-6.9) -- (+1.3,-6.5);
\filldraw[cyan] (+1.3,-6.5) circle (0.1);

\filldraw[black] (+2.85,-7.5) circle (0.1);
\draw[-] (+2.85,-7.5) -- (+2.85,-7.0);
\filldraw[cyan] (+2.85,-7.0) circle (0.1);
\draw[-] (+2.85,-6.9) -- (+2.85,-6.5);
\filldraw[cyan] (+2.85,-6.5) circle (0.1);
\draw[-] (+2.85,-6.4) -- (+2.85,-6.0);
\filldraw[cyan] (+2.85,-6.0) circle (0.1);

\filldraw[black] (+4.75,-7.5) circle (0.1);
\draw[-] (+4.75,-7.5) -- (+4.75,-7.0);
\filldraw[cyan] (+4.75,-7.0) circle (0.1);
\draw[-] (+4.75,-6.9) -- (+4.75,-6.5);
\filldraw[cyan] (+4.75,-6.5) circle (0.1);
\draw[-] (+4.75,-6.4) -- (+4.75,-6.0);
\filldraw[cyan] (+4.75,-6.0) circle (0.1);
\node [] at (+6.2,0) {1};
\draw[->,thick] (+6.2,-0.5) -- (+6.2,-1.5);
\node [right] at (+6.2,-1) {$\hat{X}$};
\node [] at (+6.2,-2) {$Z_1$};

\draw[->,thick] (+6.2,-2.5) -- (+6.2,-3.5);
\node [] at (+6.2,-4) {$Z_2$};
\node [right] at (+6.2,-3) {$\hat{X}$};

\draw[->,thick] (+6.2,-4.5) -- (+6.2,-5.5);
\node [] at (+6.2,-6) {$Z_3$};
\node [right] at (+6.2,-5) {$\hat{X}$};
\end{tikzpicture}    
\end{center}
\caption{Log sector evolution in terms of sample paths and multiplication operation $\hat{X}$ (i.e, adding a ball) in Hoppe (rooted) trees as Feynman diagrams up to order $n=3$.}
\label{fig3}
\end{figure}

\noindent The role of mutations in the one-to-one correspondence between the stochastic dynamics and the genus-zero Feynman diagrams has a natural interpretation. A mutation in the Feynman diagrams description can be thought of as the splitting of a tree and the addition of a new branch directly connected to the root of the tree, to accommodate a vertex of the new color. This is illustrated in Fig. (\ref{fig3}), with the transitions from one configuration at level $n$ to another at the next level $n+1$, labelled by the black arrows. 

\section{Markov chain and random walk on a group manifold}
Markov chains come from the study of randomly perturbed dynamical systems, and arise in various models of non-equilibrium statistical mechanics. They are therefore useful in our study of the log sector.

A Markov chain is a stochastic model that describes a process where transitions between states are governed by probability distributions. It can be defined as a sequence of random variables whose dependence on each other is characterized by the fundamental Markov property, which says that the future state of a given system is conditioned by the past only through the present state of the system. 

In the Hoppe's urn model description of the log sector, the colors of the balls in the urn after $n$ trials induce a random partition on the space $\mathcal{P}_n$ of integer partitions 

\begin{eqnarray}
\mathcal{P}_n = \left\{  \left(j_1, j_2, \ldots, j_n\right) | \left(j_1, j_2, \ldots, j_n\right) \in \mathbb{N}^n ~ \text{and} ~ j_1 + 2 j_2 + \cdots + n j_n =n  \right\}    
\end{eqnarray}

\noindent where $j_k$ is the number of colors belonging to exactly $k$ balls, $1 \leq k \leq n$. Then, according to \cite{hoppe1984polya}, the random partition is Markovian, and the distribution of the Markov process of partitions is given by the double Hurwitz numbers. In what follows, we first show how the Hurwitz numbers fit in the description of the partition-valued Markov process by demonstrating that they satisfy a recurrence relation implying the Markov property. We then show that by endowing the probability space with a group sturcture through the embedding of the Hurwitz numbers in the coproduct of the Hopf algebra of composition of functions we obtain a special type of Markov chain, a random walk on group manifolds. The Hurwitz numbers viewed as a probability distribution over the space of cycle types of permutations of $n$ objects allow to deduce that the random walk is on the symmetric group $s_n$. We finally obtain a probabilistic interpretation of the log sector of CCTMG as an $S_n$ two-dimensional gauge theory.

From the Hoppe's urn description of the log sector, we see the appearance of a random grouping process in which units (the balls in the urns) associate in (color-)clusters. We show below that the evolution of the random clustering process can be explained as a Markov process.

\subsection{Partition-valued Markov chain}
The random state of an $n$-particle in the log sector is described by the sequence $\left( j_k \right)_{k=1}^n \in \mathbb{Z}_+^\infty$, in which $j_k$ denotes the number of clusters of size $k$ satisfying $\sum_{k=1}^n k j_k=1$. Let $\left( J_k^{(n)} \right)_{k=1}^n$ be a sequence of random variables with possible realization in the sequence of values $\left( j_k \right)_{k=1}^n$. Clearly, the only possible realization of the sequence $\left( J_k^{(n)} \right)_{k=1}^n$ is the one that partitions an $n$-particle state into $j_1$ clusters of size 1, $j_2$ clusters of size 2, and so on. In other words, the sequence $\left( J_k^{(n)} \right)_{k=1}^n$ is a
random integer partition representing the cycle structure of the random permutation $\Pi^{(n)}$ of $[n]$. Equivalently, the two sequences $\left( [k]^{j_k} \right)_{k=1}^n$ in the double Hurwitz numbers of Eq. (\ref{Hurwitz numbers}) are partitions of the integer $n$. We therefore reformulate a well-known result with the original aspect being that it involves the genus zero double Hurwitz numbers with branching corresponding to two identical partitions over zero and infinity, and assert that the process $\left( J_k^{(n)} \right)_{k=1}^n$ is a Markov process with state distribution 

\begin{eqnarray}
\label{Hurwitz prob numbers}
 H^{\bullet}_{0 \xrightarrow[]{n} 0} \left\{ \left[ \left( J_k^{(n)} \right)_{k=1}^n,  \left( J_k^{(n)} \right)_{k=1}^n \right] | \left( J_k^{(n)} \right)_{k=1}^n = \left( j_k \right)_{k=1}^n \right\} = \prod_{k=1}^n \frac{1}{j_k!  (k)^{j_k}},
\end{eqnarray}

\noindent that starts at the identity partition for $n=1$, and evolves in $\mathbb{Z}^\infty_+$ with transition probabilities

\begin{subequations}
\begin{align}
H^{\bullet}_{0 \xrightarrow[]{n+1} 0} \left[ \left( j_1+1, j_2, \ldots \right),  \left( j_1+1, j_2, \ldots \right) \right] &= \frac{1}{1+n},  \label{type1}\\
H^{\bullet}_{0 \xrightarrow[]{n+1} 0} \left[ \left( j_1, \ldots, j_k-1, j_{k+1}+1, \ldots \right),  \left( j_1, \ldots, j_k-1, j_{k+1}+1, \ldots \right) \right] &= \frac{k j_k}{1+n}, \label{type2}
\end{align}    
\end{subequations}

\noindent where the transition in Eq. (\ref{type1}) occurs when $n+1$ starts a new cycle, and transition in Eq. (\ref{type2}) takes place when the element is inserted in an existing cycle of $\Pi^{(n)}$ \cite{gnedin2023random}.

To get a feeling for what the state distribution in Eq. (\ref{Hurwitz prob numbers}) says, we give a description of the profile of genus-zero Feynman diagrams in Fig. (\ref{fig3}), by looking at the variety of rooted trees and their associated sizes for small values of $n$, up to order three. Then, the question of how many tree types (a type representing an arrangement according to color clusters) are of a given size. Suppose that at level $n$, there are $J^{(n)}_1$ types represented by one tree, $J^{(n)}_2$ types represented by two trees, $J^{(n)}_3$ types represented by three trees, and so on. After $n$ steps of evolution there can be up to $n$ types, and Eq. (\ref{Hurwitz prob numbers}) specifies the probability for any $n$-tuple $\left( j_1, \ldots, j_n \right)$. This probability is 0 if the tuple is not feasible—a feasible tuple must satisfy the constraint $\sum_{k=1}^n k j_k=1$.

\noindent As an instance, refer to the second generation in Fig. (\ref{fig3}). There are two possible partitions $\left( j_1, j_2 \right)$. The leftmost configuration corresponds to the partition $j_1=2,j_2=0$, i.e to two color clusters arranged on two branches of the rooted tree, with probability $1 \times \frac{1}{2}=\frac{1}{2!}=\frac{1}{2}$. The rightmost configuration corresponds to the partition $j_1=0,j_2=1$, i.e to one color cluster arranged on the linear rooted tree, also with probability $1 \times \frac{1}{2}=\frac{1}{2!}=\frac{1}{2}$.

\noindent Let us now refer to the third generation in Fig. (\ref{fig3}). The leftmost configuration corresponds to the partition $j_1=3,j_2=0,j_3=0$,  i.e to three color clusters arranged on three branches of the rooted tree, with probability $1 \times \frac{1}{2} \times \frac{1}{3}=\frac{1}{3!}=\frac{1}{6}$. The two rightmost configurations correspond to the partition $j_1=0,j_2=0,j_3=1$, and make up one type of tree with one color cluster arranged on the linear rooted tree, with probability $2 \left( 1 \times \frac{1}{2} \times \frac{1}{3}\right)=2 \left( \frac{1}{3!} \right)=\frac{1}{3}$. Finally, the three middle configurations correspond to the partition $j_1=1,j_2=1,j_3=0$, and make up one type of tree with two color clusters arranged on two branches of the rooted tree, with probability $3 \left( 1 \times \frac{1}{2} \times \frac{1}{3}\right)=3 \left( \frac{1}{3!} \right)= \frac{1}{2}$.

\noindent The probabilities computed in the above illustration clearly appear from viewing Fig. (\ref{fig3}) as a probability tree diagram that organises the different possible outcomes of a sequence of events. Interestingly, the above arrangement is identical to enumeration problems in the case of the symmetric group $S_n$, where the conjugacy classes are indexed by partitions of $n$ corresponding to the distinct cycle types in $S_n$, and the Hurwitz number as the probability distribution in Eq. (\ref{Hurwitz prob numbers}) is the sum of all permutations in a particular conjugacy class divided by the order (i.e the number of elements) $n!$ of $S_n$.

In particular, the probability in Eq. (\ref{Hurwitz prob numbers}) given by the genus zero double Hurwitz numbers with branching corresponding to two identical partitions over zero satisfies the Markov property (also found in \cite{sibuya1993random}) 

\begin{equation}
\label{recrel}
\begin{split}
&(1+n) H^{\bullet}_{0 \xrightarrow[]{n+1} 0} \left[ \left( j_1, j_2, \ldots,j_{n+1} \right),  \left( j_1, j_2, \ldots,j_{n+1} \right) \right] = H^{\bullet}_{0 \xrightarrow[]{n} 0} \left[ \left( j_1-1, j_2, \ldots,j_n \right),  \left( j_1-1, j_2, \ldots,j_n \right) \right] \\
&+ \sum_{k=1}^n k \left( j_k + 1 \right) H^{\bullet}_{0 \xrightarrow[]{n} 0} \left[ \left( j_1, \ldots, j_k+1, j_{k+1}-1, \ldots,j_n \right),  \left( j_1, \ldots, j_k+1, j_{k+1}-1, \ldots,j_n \right) \right].
\end{split}
\end{equation}

\noindent We inductively prove Eq. (\ref{recrel}) by considering the two possible boundary cases for partitions of $n$. 

\paragraph{Case1: $j_n=1$.} In this case the partition after $n$ draws in the urn model description of the log sector must be $(j_1,j_2,\ldots,j_n)=(0,0,\ldots,1$. This case can only occur by repeating $n-1$ draws after the first one, from balls of the color as the single particle level one. In this case, we show by induction using Eq. (\ref{recrel}) that 

\begin{eqnarray}
\label{n_HN}
H^{\bullet}_{0 \xrightarrow[]{n} 0} \left[ \left( j_1, j_2, \ldots,j_n \right),  \left( j_1, j_2, \ldots,j_n \right) \right]= H^{\bullet}_{0 \xrightarrow[]{n} 0} \left[ \left( 0, 0, \ldots,1 \right),  \left( 0, 0, \ldots,1 \right) \right] = \frac{1}{n}.   
\end{eqnarray}

\noindent at levels $n=1$, $n=2$ and $n=3$, using Eq. (\ref{Hurwitz prob numbers}), we find the probabilities

\begin{eqnarray}
H^{\bullet}_{0 \xrightarrow[]{1} 0} \left[ \left( 1 \right),  \left( 1 \right) \right] = 1, \hspace{1cm}
H^{\bullet}_{0 \xrightarrow[]{2} 0} \left[ \left( 0,1 \right),  \left( 0,1 \right) \right] = \frac{1}{2}, \hspace{1cm}
H^{\bullet}_{0 \xrightarrow[]{3} 0} \left[ \left( 0,0,1 \right),  \left( 0,0,1 \right) \right] = \frac{1}{3},
\end{eqnarray}

\noindent respectively. Assume the formula holds for $n$, i.e that we have Eq. (\ref{n_HN}). We then use Eq. (\ref{recrel}) to show that this holds at the $n + 1$th generation. In that case, the partition after $n + 1$ draws must be $\left( j_1, j_2, \ldots,j_n, j_{n+1} \right)= \left( 0,0, \ldots,0,1 \right)$, and Eq. (\ref{recrel}) can be written as

\begin{equation}
\begin{split}
(1+n) H^{\bullet}_{0 \xrightarrow[]{n+1} 0} \left[ \left( 0, 0, \ldots,1 \right),  \left( 0, 0, \ldots,1 \right) \right] =& H^{\bullet}_{0 \xrightarrow[]{n} 0} \left[ \left( -1, 0, \ldots,0 \right),  \left( -1, 0, \ldots,0 \right) \right] \\
&+ 1  H^{\bullet}_{0 \xrightarrow[]{n} 0} \left[ \left( 1,-1,0, \ldots, 0 \right),  \left( 1,-1,0, \ldots, 0 \right) \right] \\
&+ 2  H^{\bullet}_{0 \xrightarrow[]{n} 0} \left[ \left( 0,1,-1,0, \ldots, 0 \right),  \left( 0,1,-1,0, \ldots, 0 \right) \right] \\
&+ 3  H^{\bullet}_{0 \xrightarrow[]{n} 0} \left[ \left( 0,0,1,-1,0, \ldots, 0 \right),  \left( 0,0,1,-1,0, \ldots, 0 \right) \right]\\
&+ \cdots \cdots \cdots \cdots \cdots \cdots \cdots \cdots \cdots \cdots \cdots \cdots\\
& + n  H^{\bullet}_{0 \xrightarrow[]{n} 0} \left[ \left( 0,0, \ldots, 1 \right),  \left( 0,0, \ldots, 1 \right) \right].
\end{split}
\end{equation}

\noindent Since the $j_k$ multiplicities cannot be negative, only the last term on the right-hand side survives. From there, plugging the right-hand side of Eq. (\ref{n_HN}) in, it is easy to see that 

\begin{eqnarray}
H^{\bullet}_{0 \xrightarrow[]{n+1} 0} \left[ \left( j_1, j_2, \ldots,j_{n+1} \right),  \left( j_1, j_2, \ldots,j_{n+1} \right) \right]= H^{\bullet}_{0 \xrightarrow[]{n} 0} \left[ \left( 0, 0, \ldots,1 \right),  \left( 0, 0, \ldots,1 \right) \right] = \frac{1}{n+1},   
\end{eqnarray}

\noindent and Eq. (\ref{recrel}) is proved in this case.

\paragraph{Case 2: $j_n=0$.} In this case, For the $n + 1$th generation, we consider a feasible partition $\left( j_1, j_2, j_n,\ldots,j_{n+1} \right)=\left( j_1, j_2,j_n \ldots,0 \right)$, and we have $1=\left( n+1 \right)^{j_{n+1}} j_{n+1}!$. Then, starting from the right-hand side (RHS) of Eq. (\ref{recrel}), we write

\begin{subequations}
\begin{align}
\begin{split}
\textrm{RHS}= & \frac{1}{1^{j_1-1} (j_1-1)! \displaystyle \prod_{k=2}^n k^{j_k}  j_k!}\\
&+ \sum_{k=1}^n k \left( j_k + 1 \right) \frac{1}{ \displaystyle \prod_{\substack{1 \leq i \leq k-1 \\ k+2 \leq i \leq n}} k^{j_k}  j_k! \cdot i^{j_i+1} (j_i+1)! \cdot (i+1)^{j_{i+1}-1} (j_{i+1}-1)!}
\end{split} 
\\
\begin{split}
=& \frac{j_1}{1^{j_1} j_1! \displaystyle \prod_{k=2}^n k^{j_k}  j_k!}\\
&+ \sum_{k=1}^n k \left( j_k + 1 \right) \frac{(i+1)j_{i+1}}{ \displaystyle \prod_{\substack{1 \leq i \leq k-1 \\ k+2 \leq i \leq n}} k^{j_k}  j_k! \cdot i^{j_i+1} (j_i+1)! \cdot (i+1)^{j_{i+1}} j_{i+1}!}
\end{split} 
\\
\begin{split}
=& \frac{j_1}{ \displaystyle \prod_{k=1}^n k^{j_k}  j_k!}+ \sum_{k=1}^n k \left( j_k + 1 \right) \frac{(k+1)j_{k+1}}{k (j_k+1) \displaystyle \prod_{k=1}^n k^{j_k}  j_k!}
\end{split} 
\\
\begin{split}
=& \frac{j_1}{ \left( n+1 \right)^{j_{n+1}} j_{n+1}!\displaystyle \prod_{k=1}^n k^{j_k}  j_k!}+ \sum_{k=1}^n  \frac{(k+1)j_{k+1}}{ \left( n+1 \right)^{j_{n+1}} j_{n+1}! \displaystyle \prod_{k=1}^n k^{j_k}  j_k!}
\end{split} 
\\
=& \left[ j_1  + \sum_{k=1}^n (k+1)j_{k+1} \right] H^{\bullet}_{0 \xrightarrow[]{n+1} 0} \left[ \left( j_1, j_2, \ldots, j_{n+1}\right),  \left( j_1, j_2, \ldots, j_{n+1}\right) \right]
\\
=& \left[ \sum_{k=0}^n (k+1)j_{k+1} \right] H^{\bullet}_{0 \xrightarrow[]{n+1} 0} \left[ \left( j_1, j_2, \ldots, j_{n+1}\right),  \left( j_1, j_2, \ldots, j_{n+1}\right) \right]
\\
=& \left( n+1 \right) H^{\bullet}_{0 \xrightarrow[]{n+1} 0} \left[ \left( j_1, j_2, \ldots, j_{n+1}\right),  \left( j_1, j_2, \ldots, j_{n+1}\right) \right]
\\
=& \textrm{LHS},
\end{align}    
\end{subequations}

\noindent which proves Eq. (\ref{recrel}) in the second case too.

\subsection{Random walk on the symmetric group}
A special class of Markov chains are random walks on groups. In this case, it is possible to understand the evolution of the log sector system in terms of probability and random walks using the underlying Hopf algebraic structure of the log partition function.

We recall that given two functions $f$ and $g$ with convergent Taylor series about $x=0$ which leave the origin invariant and are expressed as formal exponential power series

\begin{eqnarray}
f(x)= \sum_{n=1}^{\infty} \frac{f_n}{n!} x^n \quad  \text{and} \quad g(x)= \sum_{n=1}^{\infty} \frac{g_n}{n!} x^n, 
\end{eqnarray}

\noindent the composition $h=f \circ g$ is an associative operation whose Taylor series coefficients are provided by the well-known Faà di Bruno formula

\begin{eqnarray}
h_n = \sum_{k=1}^{n} \frac{f_k}{k!} \sum_j \frac{n! k!}{j_1! j_2! \cdots j_n!} \frac{g_1^{j_1}g_2^{j_2} \cdots g_n^{j_n}}{(1!)^{j_1} (2!)^{j_2} \cdots (n!)^{j_n}},   
\end{eqnarray}

\noindent where the second sum is over all $j_1,j_2, \ldots, j_n \geq 0$ such that $j_1+j_2+ \cdots+ j_n=k$. If 

\begin{eqnarray}
f(x) = \sum f_n x^n \in \mathbb{K}[[x]],~ f_0=0 ~ \text{and} ~ f_1=1,   
\end{eqnarray}

\noindent for any field $\mathbb{K}$ of characteristic zero, then $f$ has a compositional inverse $f^{-1}$, and the corresponding set of functions forms a group under composition, called the group $G^{\textrm{dif}} \left( \mathbb{K} \right)$ of diffeomorphisms tangent to the identity. The coordinate functions $a_n: f \mapsto f_n,~n \geq 1$ on $G^{\textrm{dif}} \left( \mathbb{K} \right)$ form a dual graded connected Hopf algebra called Faà di Bruno Hopf algebra, by defining a coproduct $\Delta$ on the polynomial algebra $\mathbb{K} \left[ a_1,a_2, \ldots, a_n \right]$ of coordinate functions and requiring $\Delta a_n \left( g,f \right)= a_n \left( f \circ g \right)$. The coproduct of this is commutative but not cocommutative Hopf algebra is then computed as 

\begin{eqnarray}
\Delta a_n = \sum_{k=1}^n \left( \sum_j   \frac{n!}{j_1! j_2! \cdots j_n!} \left( \frac{a_1}{1!} \right)^{j_1} \left( \frac{a_2}{2!} \right)^{j_2} \cdots \left( \frac{a_n}{n!} \right)^{j_n} \right)  \otimes a_k,   
\end{eqnarray}

\noindent and the antipode acts on each coordinate function to generate a polynomial expression for the coordinates of the compositional inverse.

Similarly, a Fa\`a di Bruno-like Hopf algebra of composition of functions can be derived as a Hopf algebra of polynomials in variables $b_1, b_2, \ldots$, where the coproduct expressed in terms of the Hurwitz numbers takes the form

\begin{eqnarray}
\label{FdB_like}
\Delta b_n = \sum_{k=1}^n \left( \sum_j   H^{\bullet}_{0 \xrightarrow[]{n} 0} \left[ \left( j_1, j_2, \ldots, j_n \right),  \left( j_1, j_2, \ldots, j_n \right) \right] b_1^{j_1} b_2^{j_2} \cdots b_n^{j_n} \right)  \otimes b_k.   
\end{eqnarray}

\noindent Any Hopf algebra $H$ equipped with a linear functional $\phi$ can be interpreted formally as leading to a generalized random walk or Markov process. In the random walk interpretation the distribution of an observable $h \in H$ after $n$ steps is given by embedding $h$ into $H^{\otimes n}$ as $\Delta^{n-1} h$ and applying the expectation value $\phi$ to each factor. The factors are the steps of the random walk and $\Delta^{n-1} h$ is understood literally as the linear combination of all the ways to arrive at $h$ via the elements in $H^{\otimes n}$ viewed as successive steps.

The comultiplication of an element in the Fa\`a di Bruno-like Hopf algebra of differential operators dual to the group associated to the log partition function plays the role of summing possibilities up according to the underlying group structure. In particular, the coproduct $\Delta b_n$ of the coordinate functions  

\begin{eqnarray}
b_n:\chi_R \mapsto \chi_{\underbrace{\Yboxdim{4pt} \yng(1) \cdots \yng(1)}_n},
\end{eqnarray}

\noindent expresses linear additions whose probabilistic interpretation is that regarding $b_n$ as random variables, $\Delta b_n$ gives the sum of possible random variables representing the system after undergoing steps in a random walk. 

A well known result of this remarkable algebraic way of understanding probability and random walks on groups is that a random walk on the Fa\`a di Bruno Hopf algebra, where the enumeration of partitions is neatly encoded in the coproduct, corresponds to a random walk on the set of partitions of the natural numbers. In our case, the Hopf algebra of differential operators generating the $\theta =1$ Hoppe trees and its associated Hoppe urns is a random walk on the symmetric group $S_n$. In what follows, we conclude with a physical interpretation of the $S_n$ random walk associated to the Hopf algebra of differential operators, for the log sector of CCTMG.

\subsection{The log sector of CCTMG as a gauge theory in two dimensions }
The equivalence between random walks on a group and two dimensional gauge theories with gauge group was pointed out in \cite{DAdda:2001yrp}, where two dimensional gauge theories of the symmetric group $S_n$ in the large $n$ limit were investigated. An important aspect of the correspondence is that the partition function of the $S_n$ gauge theory which describes branched $n$-coverings of a genus-zero Riemann surface and exhibits phase transitions can be interpreted in terms of random walks on the gauge group $S_n$. 

In our investigation of the log sector, we have shown that a similar relation holds for the log partition function mapping $\mathbb{CP}^1$ onto itself. The study of the log sector of CCTMG as a statistical model of branched covers of a Riemann surface with BEC feature being identical to a random walk on $S_n$ shows that the log sector of CCTMG is a two-dimensional gauge theory.

The use of stochastic methods to obtain the above aggregation of results has provided new directions to study the two-dimensional criticality that appears in log gravity. An interesting direction under consideration is the investigation of objects in the log sector that become random curves at the  critical point and may be characterized by stochastic dynamics of conformal maps. The theoretical framework for this study is the stochastic Loewner evolution also called Schramm Loewner evolution (SLE), a rigorous tool in mathematics \cite{schramm2000scaling,lawler2011values,lawler2011values2,lawler2003conformal} and statistical physics \cite{cardy2005sle} for generating and studying scale invariant or fractal random curves in two dimensions. Among the examples of such curves, are self-avoiding walks \cite{kennedy2002monte} and the contours of critical clusters in percolation \cite{smirnov2001critical}, Q-state Potts model \cite{rohde2011basic}, and spin glasses \cite{bernard2007possible}, as well as in turbulence \cite{bernard2006conformal}. SLE has also been applied in quantum gravity. In \cite{benincasa2017sle} for instance, the Bekenstein-Hawking entropy of a four dimensional extremal black hole was modeled in terms of a classification of particle trajectories using the framework of SLE curves. More recently, it has been argued that three-dimensional gravity can be described by Schramm Loewner Evolution (SLE) \cite{zhou2023three}. SLE is also well known to describe logarithmic conformal field theories \cite{Rasmussen:2004vb}. As our work builds up toward a physical construction for the boundary dual CFT, an interesting prospect for a new nontrivial check of the AdS$_3$/LCFT$_2$ correspondence appears. 

\section{Summary and outlook}
In this work, a further step was taken in the study of the log sector of CCTMG, thanks to a correspondence between the distribution of cluster partitions in the $n$-particle states and an allelic distribution in mathematical genetics both given by the Ewens sampling theory. The results are summarized below.

\begin{itemize}
\item We showed that the log sector can be realized as a generic one-mode urn model whose evolution is generated by adding a ball in using the multiplication generator of a Heisenberg–Weyl algebra constructed in our previous work \cite{Mvondo-She:2018htn}, in terms of differential operators acting on the polynomial components of the log partition function.
\item From the above algebraic aspect of urn model realization, we further showed that the Ewens fragmentation process encoded in the cluster-size distribution partition function $Z_{log}$ leads to a Hoppe urn model description of the log sector. 
\item The urn content of in Hoppe's model description is in one-to-one correspondence with the (genus-zero Feynman diagrams) rooted tree expansion of $Z_{log}$ found in \cite{Mvondo-She:2022jnf}, and the mutator ball in the urn model corresponds to the root in the Feynman diagram tree expansion.
\item The Markovian property of partition structures in Hoppe's work \cite{hoppe1984polya,hoppe1987sampling} was shown by the fact the genus-zero double Hurwitz numbers with branching corresponding to two identical partitions of $n$ over zero and infinity that is generated by the log partition function satisfies a recurrence relation.
\item The Hopf algebraic structure of composition of functions in $Z_{log}$ was used to show that the random dynamical system can be understood as a random walk on the symmetric group $S_n$.
\item From the fact that the partition function of any gauge theory on a two-dimensional surface of genus zero is equivalent to a random walk on the gauge group manifold, we gave a probabilistic interpretation of the log sector of CCTMG as an $S_n$ two-dimensional gauge theory.
\end{itemize}

The initial motivation that lead us to write this paper was the desire to investigate the evolution of a metastable behavior in the log sector by means of Markov chains and random walks, which constitute a crucial tool in the study of metastable dynamical systems. Metastability as a dynamical phenomenon where a multiparticle sequence of long lived unstable phases have an interesting decay dynamics generally takes place at the vicinity of a phase transition. Branch points which are singularities in the parameter space of a system at which two or more eigenvalues of a Hamiltonian and their corresponding eigenvectors coalesce and become degenerate usually play an important role in non equilibrium dynamical phase transition. In our case, the trajectories of the eigenvalues of the non-unitary Hamiltonian cross at the branch point $"\mu l=1"$, and the eigenvalues of the non-unitary Hamiltonian as well as their corresponding eigenvectors $\psi_{\mu \nu}^L$ and its log partner $\psi_{\mu \nu}^{new}$ defined as the branch point field \cite{Mvondo-She:2021joh} coalesce and degenerate. 

The log partition function describes the log sector of CCTMG as a random dynamical system of collective fields that exhibit a partition-valued Markovian, whose evolutionary stochastic behavior presented in terms of Hoppe's urn brings in the interesting result that the color labelling of the balls in the urns corresponds to a labelling by age. This parallel between age and order of appearance of novel colors in the $n$-particle evolution causes us to envisage the possibility that an LCFT dual to CCTMG could be one that takes in consideration non-equilibrium phenomena. A suggested route in the investigation of a logarithmic conformal field theory holographic to CCTMG could be to consider the counting procedure in the study of the full (single particle and multiparticle) LCFT partition function as the extension of a theory which would include a non-equilibrium Markovian dynamics. Proposals of LCFTs beyond equilibrium have recently appeared in different contexts in the literature \cite{Yashin:2022yjv,Nian:2022zum}. As it has been argued that SLE can be applied to models with non-self-crossing paths on a lattice, exhibiting self-similarity not only in equilibrium but also out of equilibrium \cite{bernard2006conformal,amoruso2006conformal,saberi2009scaling}, it would be interesting to investigate such an extension from a holographic perspective.

\paragraph{Acknowledgements} The author would like to thank the anonymous referee in the review process of \cite{Mvondo-She:2023xel} for drawing his attention to the investigation of mestability in the log sector of CCTMG. The author also wishes to thank the anonymous referee for improving the presentation of this paper. The author is grateful for the time spent at the 13th Joburg Workshop on String Theory, where this work approached its conclusion. This work is supported by the South African Research Chairs initiative of the Department of Science and Technology and the National Research Foundation, and by the National Institute for Theoretical and Computational Sciences, NRF Grant Number 65212. The support of the DSI-NRF Centre of Excellence in Mathematical and Statistical Sciences (CoE-MaSS) towards this research is also hereby acknowledged. Opinions expressed and conclusions arrived at, are those of the author and are not necessarily to be attributed to the CoE.

\clearpage

\bibliographystyle{utphys}
\bibliography{sample}
\end{document}